\documentclass[a4paper,12pt]{article}
\usepackage{amsthm,graphicx}
\usepackage{a4}
\usepackage{epsfig}
\usepackage{latexsym, amssymb} 
\usepackage{amsmath}
\usepackage{psfrag}
\usepackage{bm}
\usepackage{feynmf}
\usepackage{rotating}
\usepackage{hyperref}
\usepackage[utf8]{inputenc}
\hypersetup{
    unicode=false,          
    pdftoolbar=true,        
    pdfmenubar=true,        
    pdffitwindow=false,     
    pdfstartview={FitH},    
    pdftitle={My title},    
    pdfauthor={Author},     
    pdfsubject={Subject},   
    pdfcreator={Creator},   
    pdfproducer={Producer}, 
    pdfkeywords={keyword1} {key2} {key3}, 
    pdfnewwindow=true,      
    colorlinks=true,        
    linkcolor=red,          
    citecolor=blue,         
    filecolor=magenta,      
    urlcolor=green,         
    linktocpage=true
}

\def\beq{\begin{equation}}
\def\eeq{\end{equation}}
\def\br{\begin{eqnarray}}
\def\er{\end{eqnarray}}
\def\benu{\begin{enumerate}}
\def\efnu{\end{enumerate}}
\def\nn{\nonumber}

\def\l{\left}
\def\r{\right}

\def\cl{C_{\ell}}
\def\psk{{ P}_{k}}
\def\thefootnote{\fnsymbol{footnote}}

\begin{document}
\title{Primordial power spectrum: a complete analysis with the WMAP nine-year data}
\date{}
\maketitle
\begin{center}
\author{Dhiraj Kumar Hazra}$^{a}$\footnote{dhiraj@apctp.org}, \author{Arman Shafieloo}$^{a,b}$\footnote{arman@apctp.org} and \author{Tarun Souradeep}$^{c}$\footnote{tarun@iucaa.ernet.in}\\

{\it $^{a}$Asia Pacific Center for Theoretical Physics, Pohang, Gyeongbuk 790-784, Korea\\
$^{b}$Department of Physics, POSTECH, Pohang, Gyeongbuk 790-784, Korea\\
$^{c}$Inter-University Centre for Astronomy and Astrophysics, Post Bag 4,
Ganeshkhind, Pune 411~007, India}\\
\end{center}
\begin{abstract} 
We have improved further the error sensitive Richardson-Lucy deconvolution algorithm making it applicable directly on the 
un-binned measured angular power spectrum of Cosmic Microwave Background observations to reconstruct the form of the primordial 
power spectrum. This improvement makes the application of the method significantly more straight forward by removing some intermediate stages 
of analysis allowing a reconstruction of the primordial spectrum with higher efficiency and precision and with lower computational expenses. 
Applying the modified algorithm we fit the WMAP 9 year data using the optimized reconstructed form of the primordial spectrum with more
than 300 improvement in $\chi^2_{\rm eff}$ with respect to the best fit power-law. This is clearly beyond the reach of other alternative 
approaches and reflects the efficiency of the proposed method in the reconstruction process and allow us to look for any possible feature in the primordial spectrum projected in 
the CMB data. Though the proposed method allow us to look at various possibilities for the form of the primordial spectrum, all having good fit to the data, 
proper error-analysis is needed to test for consistency of theoretical models since, along with possible 
physical artefacts, most of the features in the reconstructed spectrum might be arising from fitting noises in the CMB data. Reconstructed error-band 
for the form of the primordial spectrum using many realizations 
of the data, all bootstrapped and based on WMAP 9 year data, shows proper consistency of power-law form of the primordial spectrum with the WMAP 9 data at all wave numbers. 
Including WMAP polarization data in to the analysis have not improved much our results due to its low quality but we expect Planck data will allow us to make a full analysis on CMB
observations on both temperature and polarization separately and in combination.  
\end{abstract}

\newpage
\tableofcontents
\renewcommand{\thefootnote}{\arabic{footnote}}
\setcounter{footnote}{0}
\section{Introduction}
Observations of Cosmic Microwave Background (CMB), in particular Wilkinson Microwave Anisotropy Probe (WMAP)~\cite{Hinshaw:2012fq} 
provide us with some of the most promising data to study the signatures
of the early universe. The initial seed of the structure formation, generated during
inflation by quantum fluctuations, leaves its imprints in the angular power spectrum in the temperature anisotropy observed in
CMB. The detected temperature and polarization angular power spectrum, contains informations 
about the background cosmology as well as the initial conditions of the universe. Precision measurements of anisotropies in the 
cosmic microwave background, and the clustering of large scale structure, suggest that the primordial density perturbation 
is dominantly adiabatic (isentropic) and has a nearly scale invariant spectrum~\cite{Hinshaw:2012fq,Hamann:2009bz,Peiris:2009wp,Samushia:2012iq,Reid:2012sw,Blake:2012pj}. This is in good agreement with most
simple inflationary scenarios which predict power law or scale invariant forms of the primordial perturbation~\cite{Starobinsky:1982ee,Guth:1982ec}. The data have also been used 
widely to put constraints on different parametric forms of primordial spectrum, mostly motivated by inflation
~\cite{ Kofman:1988xg,Salopek:1988qh,Starobinsky:1992ts,Polarski:1992dq,Adams:1997de,Chung:1999ve,Martin:1999fa,Martin:2000xs,Barriga:2000nk,Adams:2001vc,Danielsson:2002kx,Contaldi:2003zv,Elgaroy:2003hp,Kaloper:2003nv,Peiris:2003ff,Martin:2003sg,Martin:2004iv,Martin:2004yi,Hunt:2004vt,Bridges:2005br,Allahverdi:2006iq,Allahverdi:2006we,Bueno Sanchez:2006xk,Covi:2006ci,Cline:2006db,Spergel:2006hy,Hunt:2007dn,Hamann:2007pa,Joy:2007na,Joy:2008qd,Jain:2008dw,Pahud:2008ae,Flauger:2009ab,Mortonson:2009qv,Jain:2009pm,Dvorkin:2009ne,Barnaby:2009dd,Ichiki:2009xs,Hazra:2010ve,Hu:2011vr,Aich:2011qv,Hazra:2012vs}. However, 
despite the simplicity of a featureless primordial spectrum,
 it is quite important to determine the form of the primordial power spectrum directly from observations with minimal theoretical bias since departure from the scale invariant form of the PPS is motivated by most inflationary scenarios. 
 Many model independent searches have been carried out to look for features in the CMB primordial power spectrum~\cite{Hannestad:2000pm,Tegmark:2002cy,Shafieloo:2003gf,Bridle:2003sa,Mukherjee:2003ag,Hannestad:2003zs,
TocchiniValentini:2005ja,Kogo:2005qi,Leach:2005av,Shafieloo:2006hs,Shafieloo:2007tk,Nagata:2008tk,
Nagata:2008zj,Ichiki:2009zz,Shafieloo:2009cq,Paykari:2009ac,Nicholson:2009pi,Nicholson:2009zj,
bridges,Gauthier:2012aq,Hlozek:2011pc}. The Richardson-Lucy (RL) deconvolution 
 algorithm, was employed as one of the first approaches to reconstruct the form of the primordial spectrum directly from CMB data and has been extensively used till
 now on different CMB observations~\cite{Shafieloo:2003gf,Shafieloo:2006hs,Shafieloo:2007tk,Nicholson:2009pi}. In this paper we modify further the RL 
 deconvolution algorithm and we reconstruct the power spectrum using WMAP 9 year data. Amongst several new features in this paper, in particular we shall use 
 directly the {\it{un-binned}} correlated $\cl$ data to reconstruct the power spectrum. We 
note that all earlier studies have used the uncorrelated binned data to reconstruct the PPS and we show that this new modification improves the 
efficiency of the method in providing a PPS that radically improves the likelihood of fit to CMB data significantly. It is important 
to clarify that our goal is solely to determine a PPS that best fits the CMB data at any given location in the cosmological parameter space. 
We do not attempt to claim that the PPS obtained pertains to the true PPS, as such would be fraught with interpretational conundrums. Our approach 
allows a computational approach to sample the functional degree of freedom in the PPS and the consequent implications for the cosmological parameter
estimation. At first glance though it seem to be a mathematically inappropriate study, since RL needs uncorrelated data to begin
with, we shall show that the use of the un-binned data improves the algorithm in various ways for our particular case of study. We show that using this method 
for the first time we have been able to directly reconstruct a power spectrum which improves the $\chi^2_{\rm eff}$ fit to the most recent CMB data by about 200-300 better 
than the best fit power-law form of the PPS. We should mention that it is the improved quality of the present day data that allows us to directly use the un-binned
data with the modifications to the RL algorithm. This is expected to be even more relevant for the forthcoming Planck~\cite{planck} data. While getting such 
a huge improvement in the $\chi^2_{\rm eff}$ fit is exciting and hints towards the feasibility of detecting features in 
the PPS, we address the consistency check of different PPS models to the data and find that the power-law form of 
the primordial spectrum is indeed consistent with the current data after a complete error-analysis. So we reiterate to the readers that 
these two issues, reconstruction and falsification, should not be confused. The paper is organized as the following. First we discuss the RL algorithm and its modification 
implemented in this work. Then we explain in detail the way to incorporate the binned, un-binned data in a combined analysis. 
Next, we present the reconstructed results for the primordial spectrum and the corresponding angular power spectrum.
Finally we discuss the method, the results and summerize our conclusions.

\section{Formalism}~\label{sec:formalism}
In this section we shall briefly discuss the RL algorithm. We would like to
refer to the earlier papers~\cite{Shafieloo:2003gf,Shafieloo:2006hs,Shafieloo:2007tk,
Shafieloo:2009cq,Nicholson:2009pi} for detailed and more complete discussions. The method of using un-binned $\cl$ data in a combined analysis is the 
major new implication in this work and we shall explain our assumptions in subsection~\ref{subsec:unbinned}. Apart from 
obtaining a power spectrum providing a better fit, through error estimation, we address whether power law power spectrum 
is still allowed by the current WMAP data. Finally we shall aim to reconstruct PPS that provides a much better fit than the conventional 
slow roll inflationary scenarios but at the same time does 
not include too many features and can be obtained from inflationary potential with a 
limited number of parameters. 

\subsection{The Richardson-Lucy algorithm}
In the field of astronomy, RL algorithm is typically used for reconstruction of images~\cite{richardson,lucy,baugh1,baugh2}. It has been first 
demonstrated in article~\cite{Shafieloo:2003gf} that this method can be used to reconstruct the primordial power spectrum
from the angular power spectrum using the deconvolution. The observational angular power spectrum $\cl^{\rm D}$ can 
be directly related to the theoretically calculated angular power spectrum $\cl^{\rm T}$. Primordial power spectrum, 
$\psk$ generates the $\cl^{\rm T}$ by its convolution with the radiative transport kernel ${G}_{\ell k}$ through 
the following relation. 
\beq
\cl^{\rm T}=\sum_{i}{G}_{\ell k_{i}}{{P}_{k_{i}}}~\label{eq:clequation}
\eeq
It should be pointed out, barring the primordial inflationary information from $\psk$, the $\cl^{\rm T}$ {\it{only}}
depends on ${G}_{\ell k}$, which in turn is function of the cosmological parameters defining the ingredients and 
expansion rate of the universe and the history of reionization through the optical depth $\tau$ 
(for more discussion on the radiative transport kernel see ref~\cite{Shafieloo:2003gf,Shafieloo:2006hs,Shafieloo:2007tk}). 

Keeping the parameters mentioned above fixed, obtaining the $\psk$ from the $\cl$ is a deconvolution procedure. Derived 
from the elementary probability distributions, the RL deconvolution theorem iteratively solves for the primordial power
spectrum using the following algorithm. 
\beq
{{P}_{k}^{(i+1)}}-{{P}_{k}^{(i)}}={{P}_{k}^{(i)}}\times\Biggl[\sum_{\ell}
{\widetilde{G}}_{\ell k}\l(\frac{\cl^{\rm D}-\cl^{{\rm T}(i)}}{\cl^{{\rm T}(i)}}\r)\Biggr],~\label{eq:rlalgorithm}
\eeq

Here ${{P}_{k}^{(i+1)}}$ and ${{P}_{k}^{(i)}}$ are the power spectrum evaluated in iterations $i+1$ and $i$
respectively and the quantity ${\widetilde{G}}_{\ell k}$ is same as ${G}_{\ell k}$ appearing in eq.~\ref{eq:clequation}
 normalized to unity in every multipole. Here, we should point out due to noise and the cosmic variance $\cl^{D}$ has some non-zero 
error ${\sigma_{\ell}^{\rm D}}$. Since eq.~\ref{eq:rlalgorithm} does not depend on the errors associated to the 
data this method gives equal weight to every data point, which leads to a power spectrum fitting possible noise 
in the data. Hence, in this paper, we shall use and modify the Improved Richardson-Lucy (IRL) algorithm (it has first been implemented in 
ref~\cite{Shafieloo:2003gf}) which incorporates the errors in the data and in each iteration the modifications to the primordial 
power spectrum depend on the noise in the related angular power spectrum. IRL algorithm can be formulated as follows,

\beq
{{P}_{k}^{(i+1)}}-{{P}_{k}^{(i)}}={{P}_{k}^{(i)}}\times\Biggl[\sum_{\ell}
{\widetilde{G}}_{\ell k}\l(\frac{\cl^{\rm D}-\cl^{{\rm T}(i)}}{\cl^{{\rm T}(i)}}\r)~\tanh^{2}
\l[\frac{\cl^{\rm D}-\cl^{{\rm T}(i)}}{\sigma_{\ell}^{\rm D}}\r]^{2}\Biggr],~\label{eq:irlalgorithm}
\eeq
Basically the convergence factor $\tanh^{2}\l[\frac{\cl^{\rm D}-\cl^{{\rm T}(i)}}{\sigma_{\ell}^{\rm D}}\r]^{2}$, 
weighs down the contribution of two types of multipoles towards modification of the primordial spectrum, namely, 
the $\cl^{\rm T}$ which has already closer to $\cl^{\rm D}$ {\it{w.r.t}} others, and the data with higher 
errors. Here we should mention that we have scaled the data with a factor such that the primordial power 
spectrum in each iterations remains COBE normalized.


\subsection{Method of using the binned, un-binned $\cl$ in a combined analysis}~\label{subsec:unbinned}

As has been indicated in the introduction, the earlier works on the reconstruction of primordial
power spectrum have been performed using the binned data from WMAP. Although the comparison 
with the binned data seems a reasonable approximation, it has limitations. The angular power spectra 
generated from the reconstructed primordial spectra, though, gives a better
fit to the binned data compared to the power law spectrum, it provides a worse fit to the complete
data which incorporates un-binned auto and cross correlations of temperature and polarization anisotropies. As 
the reconstructed power spectrum by definition attempts to fit only the few binned data-points, in between two binned 
multipole spurious oscillations are imposed which are not expected to agree with the complete datasets and provides
a worse likelihood to the total data. Smoothing of these spurious oscillations needs to be implemented in
order to average out unwanted features and obtain a better fit to the data~\cite{Shafieloo:2003gf,Shafieloo:2006hs,Shafieloo:2007tk}.

On the other hand, though working with un-binned correlated data does not seem to be a perfect choice 
{\textit{mathematically}}, we found that this approach is much better than working with the binned data in various ways. We should remember that we are 
using IRL method adjusted for our case of study dealing with data with different uncertainties which is a clear physical problem rather than looking for 
exact solutions for a set of mathematical equations.

We can modify equation.~\ref{eq:irlalgorithm} considering the correlated error matrix of the data, henceforth we shall call it MRL (Modified Richardson-Lucy):

\beq
{{P}_{k}^{(i+1)}}-{{P}_{k}^{(i)}}={{P}_{k}^{(i)}}\times\Biggl[\sum_{\ell}
{\widetilde{G}}_{\ell k}\l(\frac{\cl^{\rm D}-\cl^{{\rm T}(i)}}{\cl^{{\rm T}(i)}}\r)~\tanh^{2}
\l[Q_{\ell} (\cl^{\rm D}-\cl^{{\rm T}(i)})\r]\Biggr],~\label{eq:modifiedirlalgorithm}
\eeq

\beq
Q_{\ell}=\sum_{\ell'}(C_{\ell'}^{\rm D}-C_{\ell'}^{{\rm T}(i)}) COV^{-1}(\ell,\ell')
\eeq

where $COV^{-1}(\ell,\ell')$ is the inverse of the error covariance matrix. 
In practice, this inverse of the error matrix from WMAP 9 years observations can not be implemented directly due to complication of likelihood estimators. However,
still one can use a likelihood function similar to the one used by WMAP team in their first data release to generate a covariance matrix using pseudo  $\cl$'s. In this 
work we show that even using the diagonal terms of the covariance matrix which are given by WMAP 9 year results would still work perfectly 
fine in our context of study. We should note here that all our results are at the end based on the likelihood codes given by WMAP mission. Looking at fig.~\ref{fig:vark} and comparing blue 
and green lines which are the likelihoods given by the WMAP codes and the $\chi^2_{\rm eff}$ derived by using the diagonal terms of the covariance matrix, we see that they are following each 
other clearly at all scales and this reflects that we have chosen a reasonable approximation in our reconstruction procedure. The new approach requires much less iterations to reach a 
reconstructed power spectra providing a better fit. On the other hand the  reconstructed power spectra directly provides a much better fit ($\sim200-300$) to the total data set 
(computed by WMAP likelihood code) compared to the best fit power law case. Finally, unlike the works with the binned datasets, 
we do not have to smooth the power spectrum in this case to achieve a good likelihood to the whole data. The use of un-binned data makes the 
method of reconstruction much more faster too, since in this case we are using more informations from the observations which directly contribute to the corrections of the PPS compared to the binned data 
where we only use 44 binned $\cl$'s which average out the contributions. Hence, in a way we get a huge improvement of fit directly with fewer iterations. Here, though to have an idealistic analysis, a treatment with full covariance matrix is required, we have found that the diagonal terms 
are sufficient to have an efficient reconstruction, at least to the cosmological scales covered by WMAP. However, we should mention that the method 
of using the un-binned data has its own disadvantages. After the multipole moment $\ell\sim900$ the quality of
data gets worse as the noise becomes comparable to the signal (or higher as we go to higher $\ell$). 
In the range $\ell=900\sim1200$ one encounters a number of negative $\cl$ which forces the MRL method 
to set $\psk$ to {\it{zero}} at high $k$ region, which is purely unphysical artefact of the negative $\cl$'s. 
To avoid this problem we have carried out our analysis by setting the negative $\cl$'s to zero. We shall demonstrate 
the results obtained in the section~\ref{sec:results}.

We find, despite of setting the negative $\cl$'s to zero (which we consider a crude approximation) the reconstructed
power spectra provide a better fit of $\sim200$ when compared with the conventional power law primordial spectrum.
We note that this is indeed an artefact of the limitations of the high $\ell$ data towards improvement of fit. Here it
is important to emphasize that as we have {\it{neglected}} the negative $\cl$'s, the amplitude of power around the third
peak is {\it{enhanced}}, which, in turn leads to an amplification of scalar power spectrum at large $k$ 
($\sim 0.1 {\rm Mpc^{-1}}-0.2 {\rm Mpc^{-1}}$). 

Finally to get rid of the {\it{unphysical}} amplification described above, we have implemented the MRL algorithm to
the combined binned and un-binned temperature data. We have considered the un-binned data till the multipole $\ell=900$
an the binned data afterwards~\footnote{We note that here too, we neglect the first negative $\cl$ at $\ell=890$ by setting it to zero.
We assume this approximation will have a negligible impact on the total likelihood}. With this procedure the 
MRL algorithm reads as,

\begin{eqnarray}
{{P}_{k}^{(i+1)}}-{{P}_{k}^{(i)}}&=&{{P}_{k}^{(i)}}\times\Biggl[\sum_{\ell=2}^{\ell=900}
{\widetilde{G}}_{\ell k}^{\rm un-binned}\Biggl\{\l(\frac{\cl^{\rm D}-\cl^{{\rm T}(i)}}{\cl^{{\rm T}(i)}}\r)~\tanh^{2}
\l[Q_{\ell} (\cl^{\rm D}-\cl^{{\rm T}(i)})\r]\Biggr\}_{\rm un-binned}\,\nn\\
&+& \sum_{\ell_{\rm binned}>900}
{\widetilde{G}}_{\ell k}^{\rm binned}\Biggl\{\l(\frac{\cl^{\rm D}-\cl^{{\rm T}(i)}}{\cl^{{\rm T}(i)}}\r)~\tanh^{2}
\l[\frac{\cl^{\rm D}-\cl^{{\rm T}(i)}}{\sigma_{\ell}^{\rm D}}\r]^{2}\Biggr\}_{\rm binned}\Biggr],~\label{eq:combined}
\end{eqnarray}

where, ${\widetilde{G}}_{\ell k}^{\rm un-binned}$ and ${\widetilde{G}}_{\ell k}^{\rm binned}$ correspond to the un-binned 
and binned kernel respectively, normalized to 1 in every multipole. 

 In an idealistic reconstruction, one should take the polarization data into account. However, the present polarization data from WMAP-9 is not good enough to improve 
 significantly our reconstruction process. For instance a quick glance of the un-binned EE angular power spectrum should reveal many negative points in every cosmological 
 scales which is certainly not physical. Binned EE and TE spectrum can be used for reconstruction as in~\cite{Nicholson:2009pi}, but we should emphasize that the huge improvement 
 of fit, which is the main crux of this paper comes from the un-binned reconstruction which certainly can not be achieved by taking into account the binned polarization data. With 
 the upcoming results from Planck~\cite{planck} we hope that 
 we shall have access to a far better polarization data which can be used for a more efficient reconstruction.


Here, we should also discuss briefly the effect of CMB gravitational lensing on the reconstruction procedure. Apart form the underlying cosmological model, the effects of lensing
on the angular power spectrum depends mildly on the form of the primordial power spectrum as well. This makes it quite complicated to incorporate the effect of lensing directly in the
RL kernel $G_{\ell k}$. However, we find that the effect of lensing for scales probed by WMAP does not differ drastically when we compare between reconstructed power spectrum and power
law spectrum. In this paper for simplicity we neglect the effect of lensing in the reconstruction process (though in the likelihood estimation it is considered) since the background model
is fixed. For the purpose of cosmological parameter estimation with the reconstructed spectra, which we are presently pursuing, we in fact take into account the effect of lensing through 
some approximation that would be reported in a forthcoming publication.



\subsection{Error-estimation}

As has been described in earlier literature~\cite{Shafieloo:2003gf,Shafieloo:2006hs,Shafieloo:2007tk}, after a few iterations, 
MRL method converges to a reconstructed power spectrum 
which provides a {\it{huge}} improvement of fit when compared with the standard power law case. It should be mentioned that, 
as the MRL method attempts to reconstruct the primordial spectra by fitting the mean values of the observed $\cl$ data, we find the 
reconstructed $\cl$ goes through the vicinity of most of the data points. Now, as the data has the probability to reside within its 
observed errors, we should be able to reconstruct a set of primordial spectra which generate $\cl$'s residing within the errors associated 
with the data. We synthesize 1000 $\cl$ datasets from the original data points with Gaussian random fluctuations with a variance associated
to the corresponding 1$\sigma$ errors to the data-points (in the context of error-estimation also see ref.~\cite{Nicholson:2009pi}). 

Using the above formalism, we generate 1000 primordial power spectra. To obtain the band associated to the uncertainty, in each mode 
we identify the most dense region which contains $68.3\%$ and $95.5\%$ (corresponding to the 1$\sigma$ and 2$\sigma$ regions respectively) 
spectra within. This statistics helps us to get rid of the distribution around the mean value and the errorband obtained is expected to 
purely contain the 1$\sigma$ and 2$\sigma$ region of maximum occurrence of events.

\subsection{Smoothing algorithm}~\label{subsec:smooth}

In this work we have used the un-binned $\cl$'s to reconstruct the primordial power spectrum. It has been indicated in earlier works
that working with the binned data requires a post-iteration smoothing~\cite{Shafieloo:2003gf,Shafieloo:2006hs,Shafieloo:2007tk} to get a power 
spectrum providing a better fit to the complete CMB data (likelihood obtained using the complete covariance matrix including 
polarization) {\it{w.r.t}} the power law spectrum. Although, the use of the un-binned data privilege us to provide a huge better fit to 
the data without smoothing, we would like to point out that this reconstructed power spectrum contains unphysical artefacts due to presence of observational and statistical noise in the data. 

Smoothing of the reconstructed power spectrum helps us to find out the broad shape 
of the power spectrum directly from the CMB observation with the averaging out the possible noise effects.

To smooth the spectrum we adopt the following algorithm using Gaussian filters. 
\beq
{{P}_{k}^{\rm Smooth}}=\frac{\sum_{{\tilde k}={\rm k_{\rm min}}}^{\rm k_{\rm max}}{{P}_{\tilde k}^{\rm Raw}
}\times\exp\l[-\l(\frac{\log{\tilde k}-\log k}{\Delta_k}\r)^2\r]}{\sum_{{\tilde k}={\rm k_{\rm min}}}^{\rm k_{\rm max}}\exp\l[-\l(\frac{\log{\tilde k}-\log k}{\Delta_k}\r)^2\r]},
\eeq

where, $\Delta_k$ indicates the width of the Gaussian filter and note that as $\Delta_k$ approaches zero, the smooth 
spectrum becomes identical to the raw one. 

In our work we have tested two types of smoothing using the filter width $\Delta_k$. In the first method we choose the 
width $\Delta_k$ to be constant over all scales. We investigate the dependence of the 
likelihood as a function of the width under different iterations. Now, keeping in mind that the quality of the data is 
not good at higher multipoles ($\ell>900$) to provide tight constraints on power spectrum at high $k$, we use a variable
width. Here, we use $\Delta_k\propto k$. This allows us to incorporate the broad oscillations (possible artifacts of 
inflationary model) at the intermediate cosmological scales and average out the violent oscillations (if any), possibly
generated from the noise. There are ways that one can define a penalty term for non-smoothness of the reconstructed 
spectrum and combine this quantity with $\chi2$ fit to the data to derive the likelihood 
(effective $\chi2$). However, the weight one gives to the penalty term and also definition
of the penalty term can both be arbitrary choices affecting the results. Though, to distinguish the broad and possible physical features from the noise, 
a careful error dependent smoothing can be carried out. We are currently investigating such issues.

\subsection{Numerical methods adopted}\label{subsec:nm}
 We have used publicly available code CAMB~\cite{Lewis:1999bs,cambsite} to compute the 
 kernel ${G}_{\ell k}$. Following the IRL algorithm
implemented in earlier works, we have developed a new FORTRAN 90 code. We have used our code as an add-on to CAMB.   
We have fixed the values of $\Omega_b,~\Omega_{\rm CDM},~H_{0}$ and $\tau$ to their best fit obtained from the analysis 
of WMAP nine-year data~\cite{Hinshaw:2012fq}. 

 For the reconstruction with the binned data we have made use of the WMAP nine-year data and its binning. We bin the 
 kernel ${G}_{\ell k}$ and the $\cl^{\rm T}$'s (appearing in eq.~\ref{eq:combined}) using the same binning as 
 in WMAP nine-year data. We normalize ${G}_{\ell k}$ to one in every multipole as indicated in earlier works
 to get ${\widetilde{G}}_{\ell k}$. For the analysis with the un-binned data we follow the same procedure as above 
 except the binning and for the combined data we follow eq.~\ref{eq:combined} and use the binned and the un-binned data 
 as required by the algorithm. For the initial guess we have used the best fit primordial power spectrum from WMAP-9. 
 We have used nearly 1700 $k$-space samples to calculate the $\cl$'s. Using a higher number of $k$-space points 
 obviously provides a much better fit to the spectra ($\Delta\chi^2_{\rm eff} > 300$) as it contains more degrees of freedom, but
 at the same time contains violent oscillations at higher wavenumbers. If we aim to converge on a power spectrum that 
 can be motivated by an inflationary theory with a few parameters, we should have less fluctuations in the power 
 spectrum and it requires a careful smoothing of the spectra which is of course easier to implement in the former case 
 with less $k$-space sampling.

\section{Results}\label{sec:results}
\begin{figure}[!htb]
\begin{center} 
\psfrag{chi2}[0][1][1.2]{$-\ln{\cal L}$} 
\psfrag{variance}[0][2][1.2]{$\Delta_{k}$}
\resizebox{400pt}{270pt}{\includegraphics{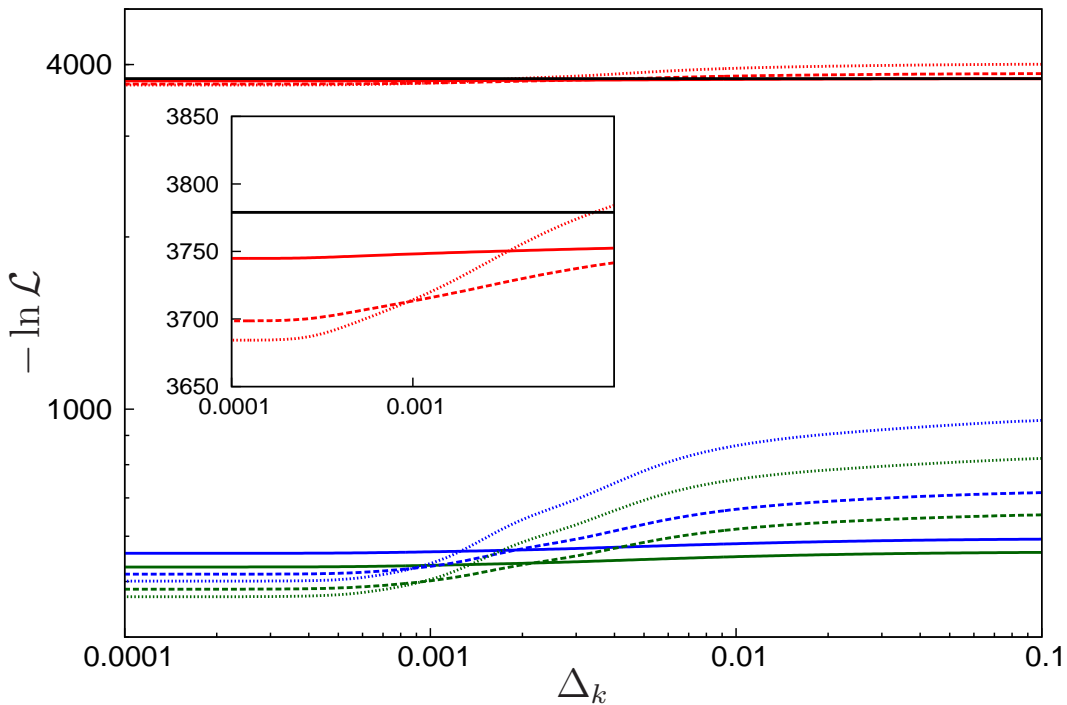}} 
\end{center}
\caption{\footnotesize\label{fig:vark} The dependence of the likelihood on the width 
of the smoothing filter in logarithmic scale. In red we have plotted the $-\ln{\cal L}$ of the total data
obtained using the WMAP likelihood. Using the diagonal term of the inverse covariance matrix
the quality $-\ln{\cal L}$ is plotted using simple $\chi^{2}$ statistic (in green) and using 
the WMAP likelihood (in blue) for the temperature auto-correlation data only.In black we have 
indicated the total $-\ln{\cal L}$ in the case of power law. Moreover here we illustrate the results
of the smoothing after three different iterations. Results for iterations 100 (in dots), 70 (in dashed-lines) and 20 
(in solid lines) are shown. The inset provides the total $-\ln{\cal L}$ for three different iterations in linear scale.
The dotted red line (for 100 iterations) suggests that we are able to achieve a better fit of about 200 {\it{w.r.t.}} the 
power law likelihood.}
\end{figure}

In this section we shall demonstrate the results of our analysis. We should point out that, unless otherwise explicitly mentioned,  
the figures presented in this section are obtained using the combined un-binned and binned data following eq.~\ref{eq:combined}. 
Following the smoothing algorithm in the subsection~\ref{subsec:smooth} we first illustrate the effects of a constant smoothing width
$\Delta_k$ on the likelihood (${\cal L}$). In Fig.~\ref{fig:vark} we have shown the quantity $-\ln{\cal L}$
 (or $\chi^2_{\rm eff}/2$)as a function of the smoothing width. This figure shows the likelihood calculated considering temperature 
data alone and including the polarization. We would like to point out that smoothing of the primordial spectrum 
affects the temperature and the polarization spectrum in a similar way and this fact justifies that our method of working with the 
un-binned data is quite apt. Further the inset of the figure suggests using the combined un-binned and binned data only after 20
iterations we are able to get a better fit of $\sim70-80$ without any smoothing of the data~\footnote{Note that the inset 
captures the part of the figure where the smoothing width negligible and as it has been discussed in the preceding section
the smoothed spectrum matches the actual spectrum for low value of $\Delta_k$}. Needless to add, as expected we get better 
fit with the low width of the smoothing filter. However, we would like to add a couple of sentence regarding the 
increase of $\chi^2_{\rm eff}$ obtained from highly smoothed spectra as we increase iterations. A quick 
glance at the WMAP un-binned data reveals for a number of multipoles the theoretically angular power spectrum 
from power law is unable to fit the data. 
Basically the $\tanh$ convergence factor introduced in the IRL (and MRL) algorithm induces the broad features in the 
spectrum which helps us to fit the the outliers with low noise in first few iterations. If we allow higher iterations
the it is possible to get a huge improvement of fit ($\sim~300$) by fitting the possible noise in the data. Now 
a smoothing of such highly oscillatory spectrum using a filter of uniform width in all scales suppresses the imposed 
oscillations in uniform manner which leads to a worse fit. Fig.~\ref{fig:vark} suggests that for $\Delta_k\sim0.1$ 
the spectra obtained after 70 and 100 iterations actually fit the data worse than the power law, while note that 
the smoothed spectra after 20 iterations is indeed providing a better fit. However, with decreasing $\Delta_k$
we recover that with higher iterations we get higher better fit to the data. Here, we should also mention that 
using a different smoothing algorithm it should be possible to get rid of unwanted oscillations in the spectrum. 
We find smoothing filter with width proportional to the wavenumbers can be more effective 
in providing a power spectrum containing only large oscillations (expected to be physical artefacts) which also 
provides a reasonable better fit to the data. 


\begin{figure}[!htb]
\begin{center} 
\psfrag{psk}[0][1][1]{${P}_{k}$} 
\psfrag{k}[0][1][1]{$k~{\rm Mpc^{-1}}$}
\psfrag{ell}[0][1][1.2]{$\ell$}
\psfrag{ll1cltt}[0][1][1]{${\ell(\ell+1)\cl^{\rm TT}}/{2\pi} \l[\mu{\rm K}^{2}\r]$}
\psfrag{ll1clee}[0][1][1.2]{${\ell(\ell+1)\cl^{\rm EE}}/{2\pi} \l[\mu{\rm K}^{2}\r]$}
\psfrag{l1clte}[0][1][1.2]{${(\ell+1)\cl^{\rm TE}}/{2\pi} \l[\mu{\rm K}^{2}\r]$}

\resizebox{350pt}{240pt}{\includegraphics{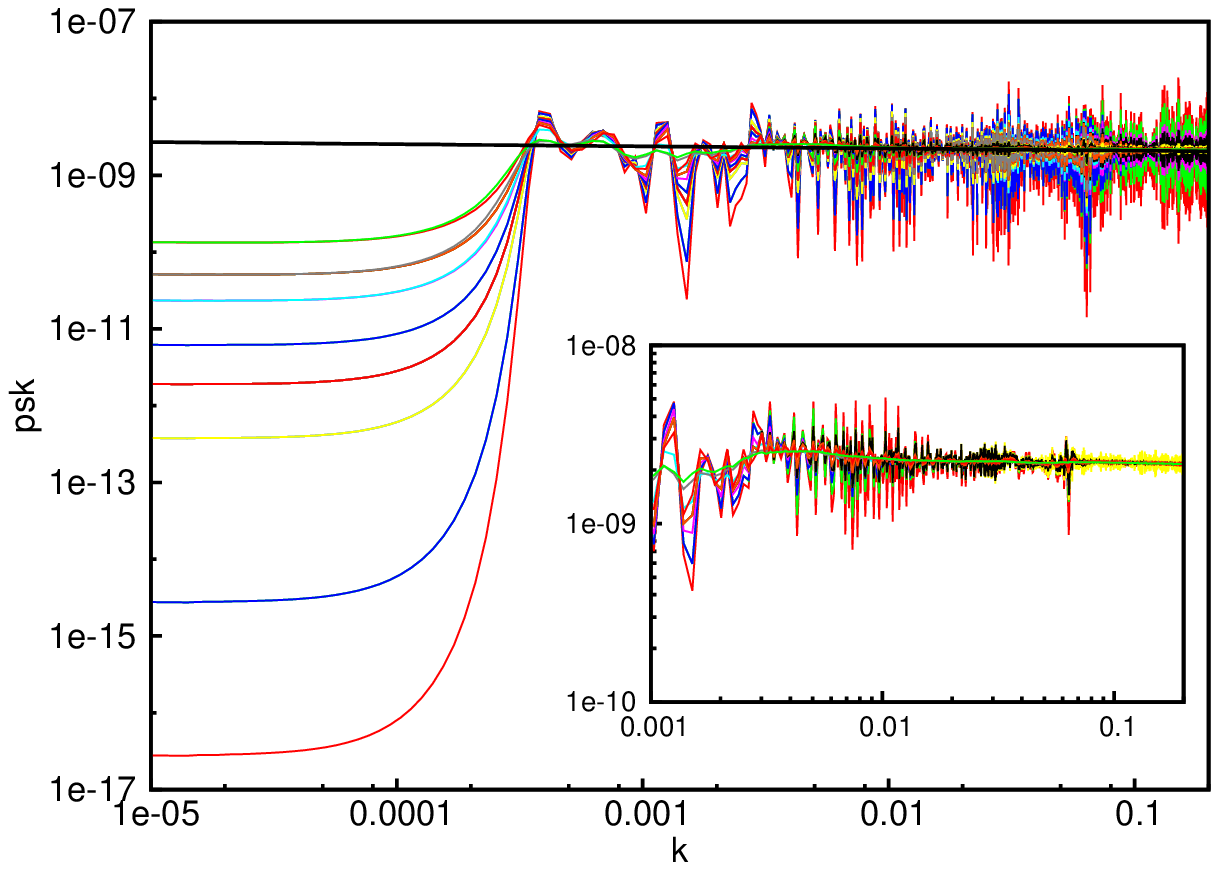}} 
\resizebox{350pt}{240pt}{\includegraphics{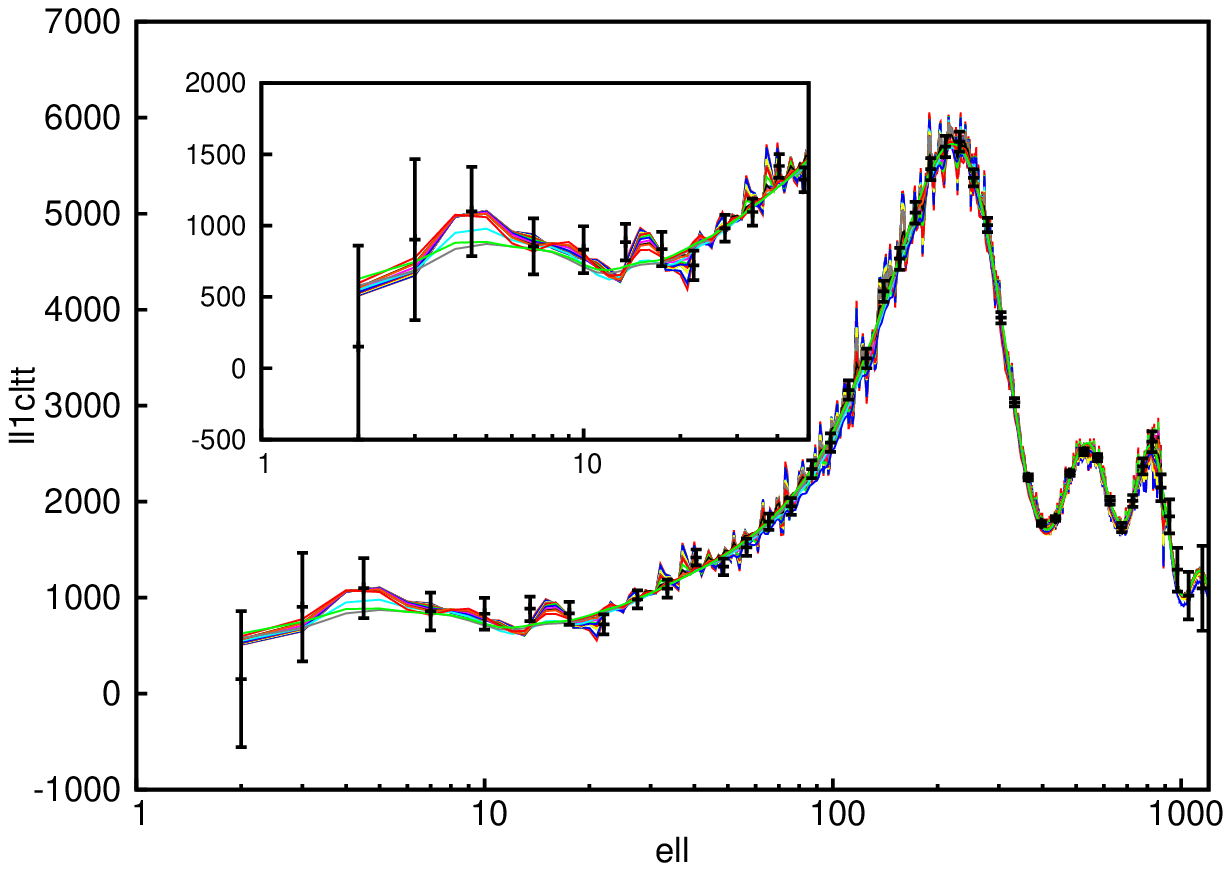}} 

\end{center}
\caption{\footnotesize\label{fig:sampleplots} The reconstructed primordial power-spectra (on top) and the corresponding angular 
power spectra (at the bottom) that provide a better fit ranging from $\sim2-300$ compared to the power law model. We have plotted 
20 sample power spectra (for different iterations and different smoothing width) and the corresponding angular power spectrum along 
with the data with error-bars (in black). As expected the reconstructed power spectrum can address all the outliers that can not be 
fit be the power law model. The inset on the top figure contains 10 sample power spectrum where a using a high width of smoothing
we have been able to average out the oscillations responsible for fitting the possible noise in the data. Note that at high k the 
smooth spectrum is nearly scale invariant and is in good agreement with the power law (in black). The Sachs-Wolfe Plateau region
is highlighted in the inset of the second figure which illustrates the fitting of the low-$\ell$ outliers by the features in the 
reconstructed power spectrum. 
}
\end{figure}
\begin{figure}[!htb]
\begin{center} 
\psfrag{psk}[0][1][1]{${P}_{k}$} 
\psfrag{k}[0][1][1]{$k~{\rm Mpc^{-1}}$}
\psfrag{ell}[0][1][1.2]{$\ell$}
\psfrag{ll1cltt}[0][1][1]{${\ell(\ell+1)\cl^{\rm TT}}/{2\pi} \l[\mu{\rm K}^{2}\r]$}
\psfrag{ll1clee}[0][1][1.2]{${\ell(\ell+1)\cl^{\rm EE}}/{2\pi} \l[\mu{\rm K}^{2}\r]$}
\psfrag{l1clte}[0][1][1.2]{${(\ell+1)\cl^{\rm TE}}/{2\pi} \l[\mu{\rm K}^{2}\r]$}

\resizebox{180pt}{130pt}{\includegraphics{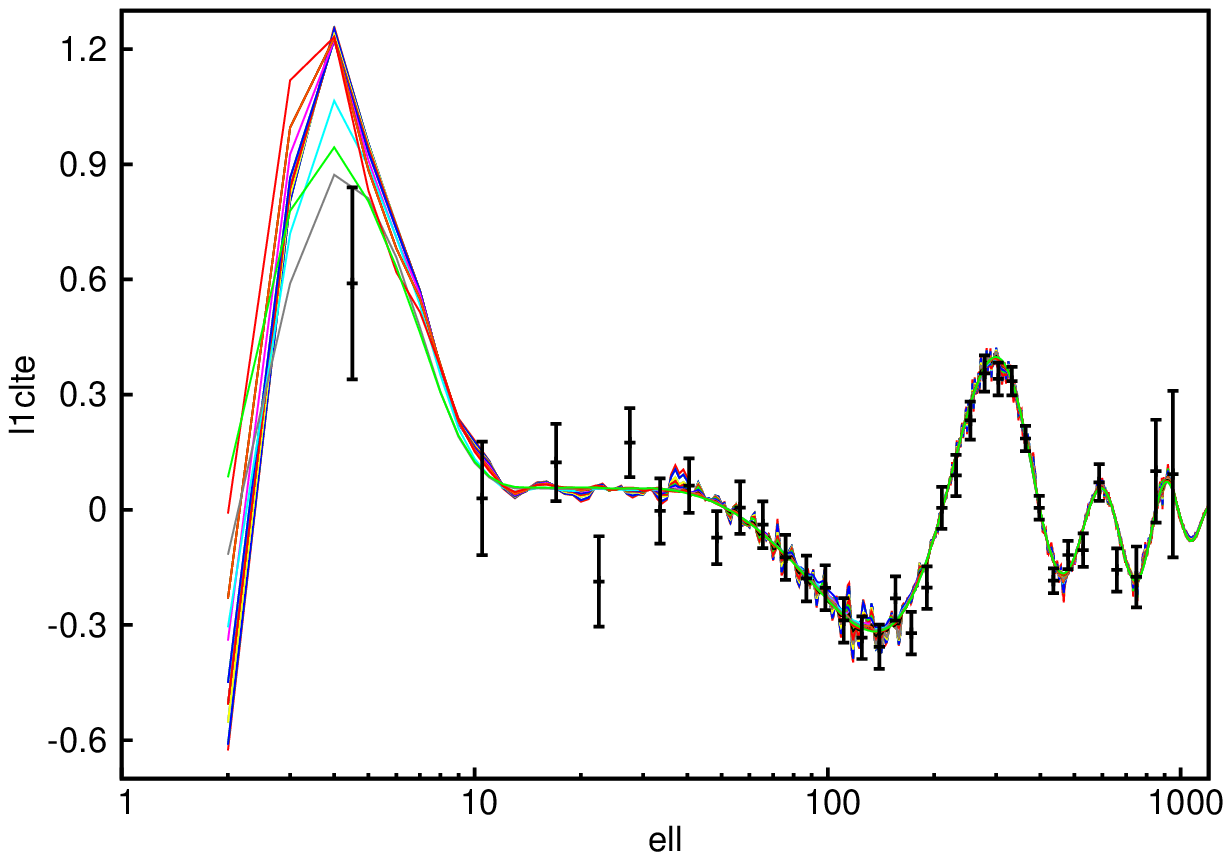}} 
\resizebox{180pt}{130pt}{\includegraphics{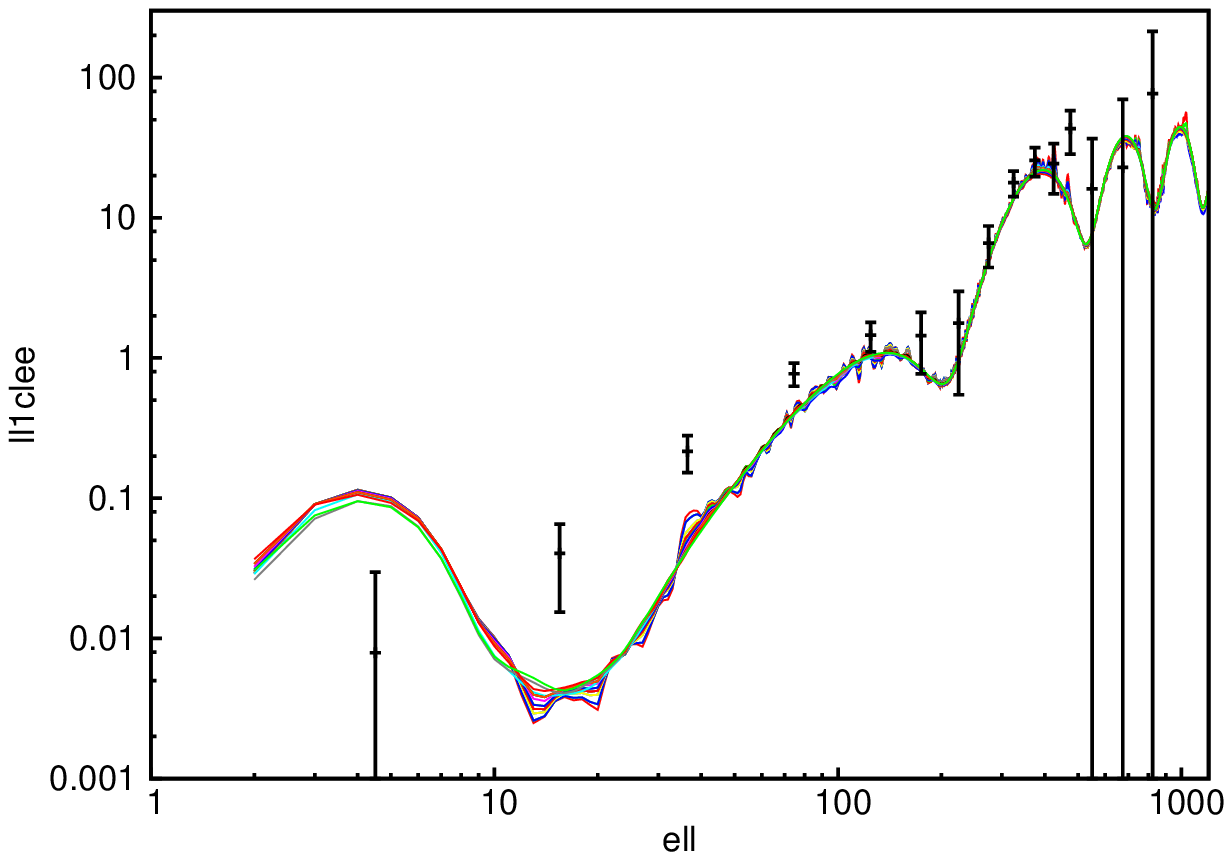}} 

\end{center}
\caption{\footnotesize\label{fig:sampleplotsteee} The ${C}_{\ell}^{\rm TE}$ and ${C}_{\ell}^{\rm EE}$ obtained from the 
20 reconstructed power spectrum as shown in the fig.~\ref{fig:sampleplots}. The data and the error bars are plotted in black. It 
should be highlighted that while the theoretical $\cl$'s are largely in agreement with the TE data, for EE data there are a few 
outliers as well.  
}
\end{figure}

A few reconstructed primordial power spectra and the corresponding ${C}_{\ell}^{\rm TT}$ are plotted in 
fig.~\ref{fig:sampleplots}. ${C}_{\ell}^{\rm TE}$ and ${C}_{\ell}^{\rm EE}$ associated with the same set 
of plots are given in fig.~\ref{fig:sampleplotsteee}. In both the plots black dots with error-bars are the data points
while the black straight line on fig.~\ref{fig:sampleplots} refers to the best fit primordial power spectrum from WMAP-9.
Noticeably, the plot suggests that there are no '{\it{outliers}}' to the reconstructed spectrum. 
To arrive at these figures we have used the MRL algorithm with the combined un-binned and binned data. As it has been indicated 
in the formalism section, due to the bad quality of the data at high $\ell$ ranges, we encounter several negative $\cl$'s in high multipoles. In fact since the radiative transport kernel ${G}_{\ell k}$ is a positive definite matrix and $\psk$ are positive, the $\cl$'s also should be positive. We have found that MRL method forces the $\psk$ to be zero to high wavenumbers to match the
unphysical and negative data points evident leads to wrong reconstruction. To avoid the negative $\cl$'s we have imposed
zero values to the $\cl$'s to the multipoles where the data goes negative. While this assumption stops $\psk$ to go to 
zero at at high $k$, we get an overall amplification of power at high $k$ as imposing the negative $\cl$'s to zero 
introduces an average enhancement of power at high $\ell$'s. However, we would like to mention that, despite of this
amplification we get a better fit of about $\sim200$ as the high $\ell$ data has relatively low contribution on the 
overall likelihood. The combination of un-binned and the binned data on the other hand works better in this cases. 
As described in eq.~\ref{eq:combined} we have used the un-binned data till $\ell=900$ and binned data thereafter. 
We encounter only one negative $\cl$'s at $\ell=900$ and we set it to zero as has been indicated earlier. We find that
this method is the optimum choice as the reconstructed $\psk$ using combined data provides a better likelihood 
compared to the $\psk$ obtained using the binned data and the power spectrum is completely free from the unphysical 
amplification at high $k$.

\begin{figure}[!htb]
\begin{center} 
\psfrag{psk}[0][1][1.2]{${P}_{k}$} 
\psfrag{k}[0][2][1.2]{$k~{\rm Mpc^{-1}}$}
\resizebox{200pt}{135pt}{\includegraphics{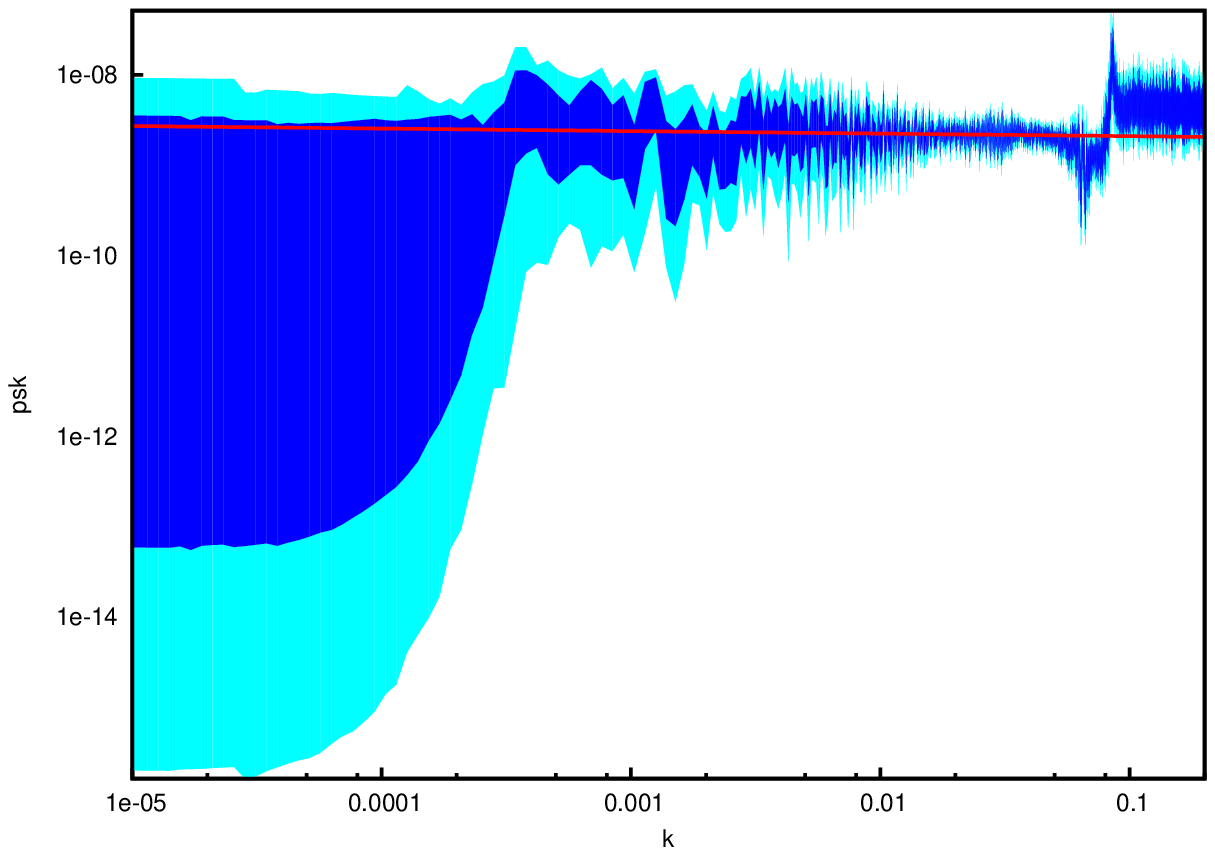}} 
\resizebox{200pt}{135pt}{\includegraphics{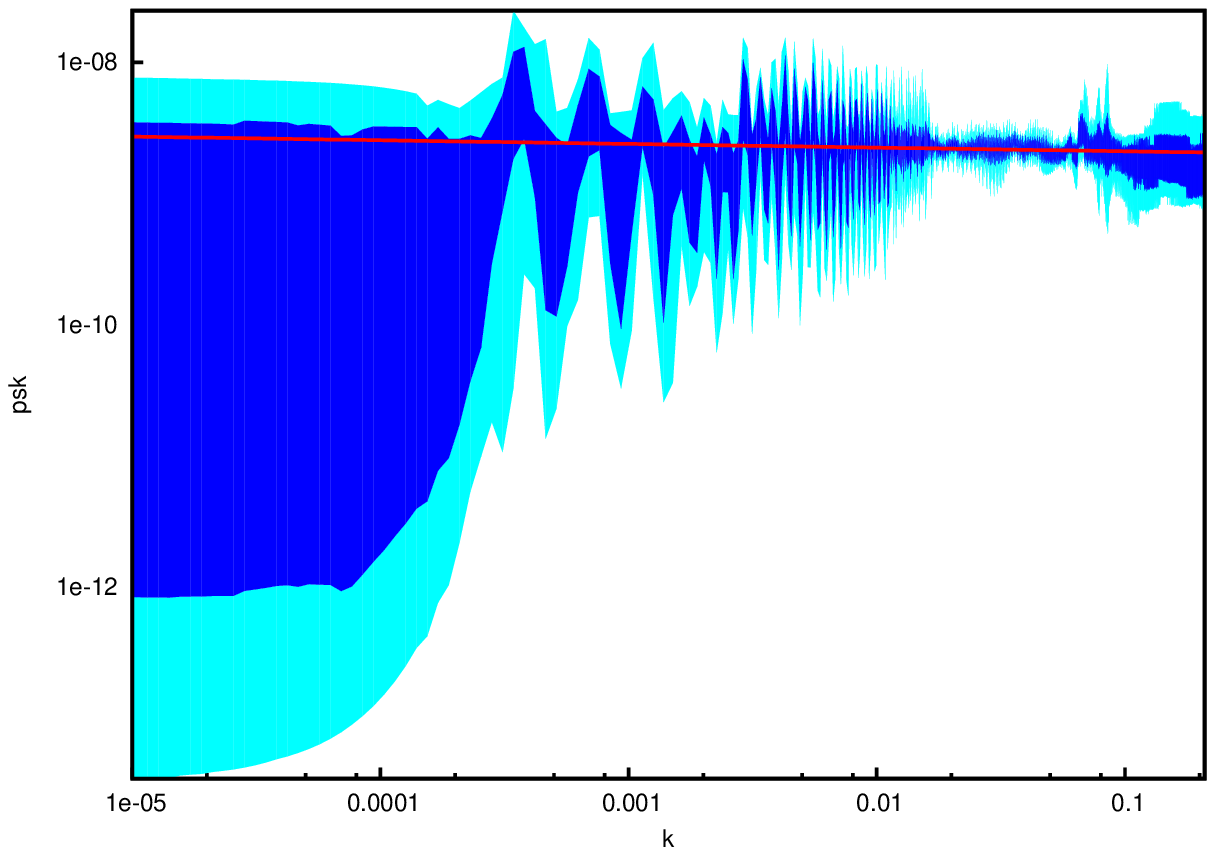}} 
\end{center}
\caption{\footnotesize\label{fig:ps-error} The 1-$\sigma$ (blue) and 2-$\sigma$ (cyan) estimated error-band associated to 
reconstructed primordial power spectrum. On the left we have plotted the results obtained from the analysis of the un-binned data 
only and on to the right we have plotted the corresponding results from the binned reconstruction. The red line is the best fit 
power law primordial power spectrum from WMAP-9.
}
\end{figure}

\begin{figure}[!htb]
\begin{center} 
\psfrag{psk}[0][1][1.2]{${P}_{k}$} 
\psfrag{k}[0][2][1.2]{$k~{\rm Mpc^{-1}}$}
\resizebox{400pt}{270pt}{\includegraphics{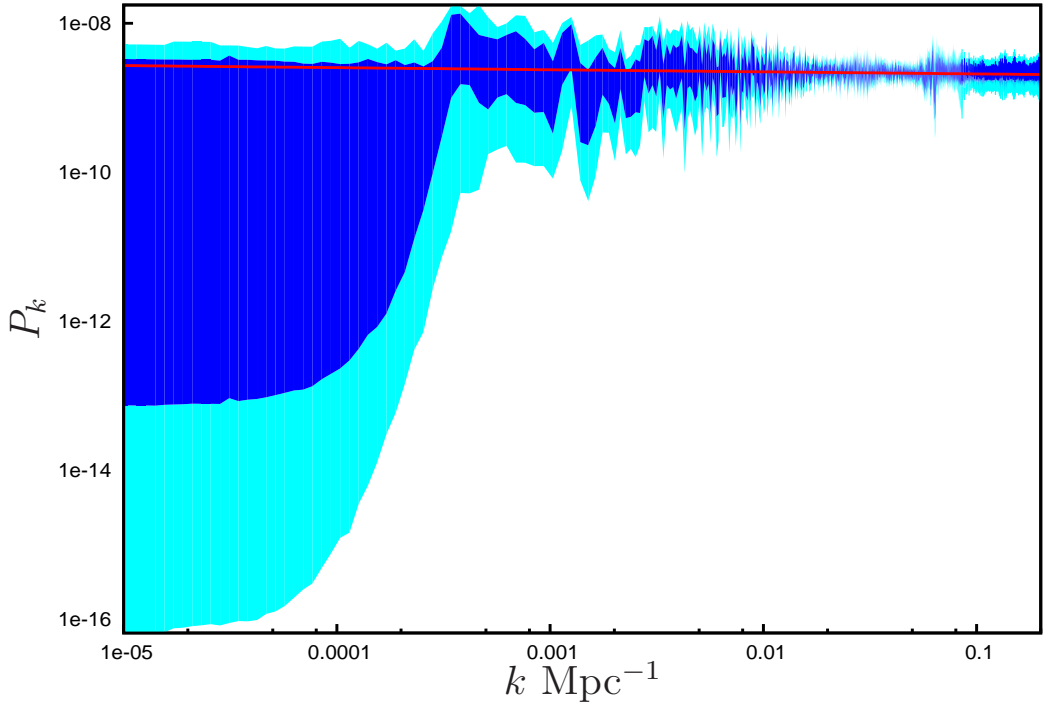}} 
\end{center}
\caption{\footnotesize\label{fig:combinedps-error} The 1-$\sigma$ (blue) and 2-$\sigma$ (cyan) estimated error-band associated to 
reconstructed primordial power spectrum obtained using the combined data, which we find to be the optimum choice for the power 
spectrum reconstruction. Like the previous figure, the red line stands for the best fit power law primordial spectrum, indicating 
that it is still well within the above said 1$\sigma$ errorband.    
}
\end{figure}

The error estimation is an integrated part of the study of the power spectrum reconstruction. We have shown that the 
MRL method using the the combined data can reconstruct a power spectrum capable of providing a huge fit to the data. 
However only reconstruction of a spectrum does not guarantee its evidence. It is expected that the observed data is likely to
reside within its errors. At this stage it is important to address what is the range of uncertainty in the primordial power 
spectrum corresponding to the error bars associated to the observed data and whether the power law power spectrum falls
within the uncertainty band. As discussed in the previous section, we have generated 1000 realizations from the actual data 
using random noise with Gaussian distributions of variance associated to the errors of the data. Using MRL algorithm we find 1000
reconstructed power spectrum from which, in each $k$, we have chosen the most concentrated region containing $68.3\%$ and $95.5\%$
spectra corresponding to 1$\sigma$ and $2\sigma$ error bands respectively. We do not use distributions around the mean value 
as we think it might be biased towards the spectrum providing better fit only. We have implemented this statistic to obtain 
the error bands for the case of reconstruction with the un-binned and binned data (see, fig.\ref{fig:ps-error}) and the 
combined data (see, fig.\ref{fig:combinedps-error}). These 3 figures contain the 1$\sigma$ and 2$\sigma$ errors associated 
to the power spectra indicated by blue and cyan region respectively. Notice that the left panel of fig.~\ref{fig:ps-error}, which 
illustrates the result of the un-binned data indicates an overall enhancement of power at high $k$ values ($k\sim0.1~{\rm Mpc}^{-1}$).
We would like to mention that this is simply the artifact of imposing zero values for the negative $\cl$'s. However, the reconstruction
with the binned data does not contain any such amplifications for obvious reasons. It should be emphasized once again that 
binned reconstruction does not {\it{directly}} provide a better fit to the complete dataset.  

The red line in the 3 plots in figures~\ref{fig:ps-error} and~\ref{fig:combinedps-error}
describes the best fit power law primordial spectrum. As it is evident from these 3 figures, we would like to report that in all the 
cases we find that power law spectrum lies {\it very much} within the $1\sigma$ errors of the data. 
Now, as we have argued, we find our result using the combined data is free from the unwanted low scale amplification of power and it directly
provides a better likelihood to the total data. It is expected with the upcoming results from Planck we shall have access to tightly 
constrained $\cl$'s. It will be exciting to examine whether we can achieve a highly constrained primordial power spectrum using similar formalism. 
Further as Planck data will provide the $\cl$'s till $\ell\sim2000$ (considerably higher than WMAP), it will certainly allow the reconstruction 
for wider cosmological scales.

 Finally we should mention that, we have incorporated the polarization data  by considering the 
combination of $C_{\ell}^{\rm TT}+2C_{\ell}^{\rm TE}+C_{\ell}^{\rm EE}$ following the similar reconstruction procedure. We find the reconstructed power spectrum
is almost similar to the spectrum reconstructed from the TT data alone without significant improvement in the fit which simply reflects the low quality of the current polarization data for the purpose of PPS reconstruction.

\section{Discussion}\label{sec:discussion}
In this paper we have made a new modification to the Richardson-Lucy deconvolution algorithm~\cite{Shafieloo:2003gf,Shafieloo:2006hs,Shafieloo:2007tk,Shafieloo:2009cq,Souradeep:2008zz} by using the whole un-binned 
correlated CMB data directly in the reconstruction of the primordial power spectrum. We have shown that the new modification can improve 
the efficiency of the method significantly in gaining better fit to the data and reducing the computational expenses. We have applied the 
method on the most recent CMB data from WMAP 9 years observations and we managed to get a better fit of more than $300$ in $\chi^2_{\rm eff}$ with respect 
to the best fit power-law. We have demonstrated, that unlike the binned reconstruction, to get an improvement of fit to the whole CMB data using the likelihood 
code provided by WMAP team we do not have to smooth the reconstructed results. However, due to the presence of some negative and unphysical $C_l$’s in the 
un-binned data the reconstructed power spectra have been contaminated with an unphysical enhancement in power at small scales. This is due to the fact that RL 
method is applicable only on positive definite matrices and both $C_l$ and $P(k)$ should be positive. We have solved this enhancement problem by combining the 
un-binned and binned data. We have used the un-binned data till the multipole where the quality of the data is good, basically for $l<900$ and have used the binned 
data for the multipoles after it. We have shown that using the MRL (Modified Richardson-Lucy) algorithm with the combined data is an optimum choice for reconstruction of the primordial spectrum. 
We have also performed smoothing on the reconstructed results to get variety of the cases with different fluctuations that could rise to the high likelihood to the data.
We have presented a few samples of reconstructed smooth primordial power spectrum and the corresponding angular power spectra which provide better likelihoods w.r.t. the 
power law spectrum. Here we should also mention that we have implemented the smoothing to obtain a primordial power spectrum that can be generated with only a handful of parameters such 
that this can be motivated from an inflationary scenarios. After comparing the effect of the width of smoothing filter  on the total likelihood and the likelihood obtained from the diagonal 
term of the inverse covariance matrix we argue that our method of using the combined data (un-binned and binned) in MRL algorithm is a robust assumption. Finally to obtain the error band for the 
reconstructed  primordial spectrum associated to the uncertainty of the observational data we generated 1000 datasets bootstrapped from the original $C_l$’s.  We find that though the features 
provide a huge better fit to the data, all these features could be presented because of random fluctuations and noise and in fact power law is still very much consistent to the CMB observations. 
We should emphasis here that there is a delicate difference between reconstruction and falsification. While doing falsification we test possibility of the observed data given the model, in the reconstruction approach we try to look at all 
phenomenological possibilities. In this work we have done both of these tasks and while we see that power-law is consistent to the data, we present handful of cases and forms for the 
primordial spectrum that all can give very good fit to the data. With the upcoming data from Planck we expect that it will be possible to use un-binned data up to higher range of multiples
and with our proposed procedure we will be able to probe smaller scales of  primordial power spectrum. Further, we also expect to get a tighter errorband on the primordial spectrum than what 
we have achieved with WMAP-9 using high quality of Planck polarization data.  In fact the quality  of the WMAP polarization data was very much inferior to its temperature data and we could not 
achieve any significant better fit to the whole data by incorporation of WMAP polarization data. While in this paper we proposed a new and easy to use approach for the reconstruction of the primordial 
spectrum, in a companion paper we perform the important task of 
cosmological parameter estimation allowing the free form of the PPS~\cite{cpeffps}.


\section*{Acknowledgments}
D.K.H and A.S wish to acknowledge support from the Korea Ministry of Education, Science
and Technology, Gyeongsangbuk-Do and Pohang City for Independent Junior Research Groups at 
the Asia Pacific Center for Theoretical Physics. We also acknowledge the use of publicly 
available CAMB to calculate the radiative transport kernel and the angular power spectra. 


\end{document}